# Josephson spectroscopy study of kagome superconductors toward the deep point-contact regime

Hailang Qin[1†], Xiao-Yu Yan[2], Hanbin Deng[2], Mu-Wei Gao[2], Guowei Liu[1], Yuanyuan Zhao[1], Jia-Xin Yin[1,2‡]

[1]Quantum Science Center of Guangdong-Hong Kong-Macao Greater Bay Area, Shenzhen, China.

[2]State Key Laboratory of Quantum Functional Materials, Department of Physics, and Guangdong Basic Research Center of Excellence for Quantum Science, Southern University of Science and Technology, Shenzhen 518055, China

[†]Contact author: qinhailang@quantumsc.cn

[‡]Contact author: yinjx@sustech.edu.cn

**ABSTRACT**. Josephson scanning tunneling microscopy (JSTM) has emerged as an important technique for probing the superconducting order parameter at the atomic scale. However, the Josephson current in JSTM may behave quite differently when the coupling strength varies. Here, we push the junction to the deep point-contact regime, reaching a normal-state junction resistance of only $0.15 \cdot h/2e^2 \approx 2\ k\Omega$. We demonstrate, using kagome superconductors, that the zero-bias conductance, a key characteristic of the Josephson current, deviates strongly from the quadratic dependence on the normal-state conductance upon entering the deep point-contact regime. Furthermore, we observe a striking saturation of the zero-bias conductance, which we show arises from the series resistance in the circuit. This also serves as a cautious reminder when interpreting zero-bias conductance saturation or quantization in studies of exotic physics such as that of Majorana zero modes if the tip-sample junction resistance is extremely small. Finally, we identify an optimum regime where JSTM can be used as an atomic-scale probe for studying pair-density wave states in materials with low superconducting transition temperature, such as $AV_3Sb_5$ kagome superconductors.

## I. INTRODUCTION

The direct-current (DC) Josephson effect [1-3] is the phenomenon of a supercurrent due to Cooper-pair tunneling, flowing between two superconductors separated by a thin insulating layer when there is no voltage across the junction. It was first theoretically predicted by Josephson [1] and experimentally observed by Anderson and Rowell [4]. The supercurrent can be expressed as $I_J = I_C \sin\varphi$, where $\varphi$ is the phase difference between the two superconductors and $I_C$ is the critical current, i.e., the maximum current that can be applied before a voltage appears across the junction [2, 3, 5].

The Josephson effect can be exploited in a scanning tunneling microscope (STM) by using a superconducting tip and a superconducting sample, effectively forming an ultrasmall superconductor-insulator-superconductor (SIS) Josephson junction, a technique termed JSTM [6-9]. JSTM has since been employed to investigate various properties of superconductors, such as vortices, order parameter variations, and more recently the pair-density wave (PDW) states in cuprates and $AV_3Sb_5$ kagome superconductors [8, 10-20]. The major features of the current-voltage (I-V) characteristics of a typical JSTM junction include, in addition to the single-particle tunneling features discussed below, the appearance of Cooper-pair supercurrent near zero bias. The current increases gradually from zero bias and reaches a maximum $I_{max}$, before decreasing. This maximum current is usually directly related to the critical current $I_C$ of the junction and can be used to extract the superconducting order parameter of the sample and study the homogeneity of the order parameter on the nanometer and even atomic scale [12, 21]. Correspondingly, a zero-bias peak (ZBP) appears in the dI/dV conductance spectrum owing to the supercurrent. These common features in the I-V and dI/dV spectra are usually understood using the phase-diffusion model [22] or the P(E) theory in the dynamical Coulomb blockade regime, where the Cooper pairs tunnel inelastically by exchanging energy with the electromagnetic environment [21, 23-25].

As the tip-sample distance decreases, which corresponds to the decrease of the normal resistance of the junction, the ZBP and the maximum supercurrent generally increase. This can be qualitatively understood, considering that the Josephson coupling between the superconducting tip and superconducting sample increases. Naturally, one would wonder what would happen if the coupling becomes even stronger, strong enough that the tip-sample junction resistance reaches the resistance quantum $R_0 = h/2e^2 \approx 12.9\ k\Omega$ or even smaller. Experimentally, this can be difficult, as the junction may become unstable and the tip-sample junction transitions from tunneling to the point-contact regime.

Here, we report a comprehensive study of the behavior of the Josephson current and the corresponding zero-bias conductance (ZBC) when the tip-sample junction normal-state conductance $G_n$, defined as the setpoint current divided by the voltage across the junction (which is less than the bias setpoint due to the voltage drop across the series resistance in the circuit), crosses from the tunneling regime to the point-contact regime. As $G_n$ increases, we observe a clear transition from the tunneling regime to a quantum point-contact. The ZBC increases approximately quadratically with $G_n$ even in the tunneling-to-point-contact crossover regime. After entering the point-contact regime, ZBC increases much faster, likely due to multiple Andreev reflections. The ZBC finally appears to saturate, which is found to be mainly due to cable resistance in series with the junction in the circuit. With a comprehensive understanding of the Josephson current in JSTM in both tunneling and point-contact regimes, we further identify the optimal condition for JSTM to be used as an atomic-scale probe for novel quantum states, like PDW.

## II. METHODS

The experiments were performed using an STM equipped with a dilution refrigerator, at a base temperature of about 30 mK [16]. We study two types of samples in this work. The first sample is a single crystal of $Cs(V_{0.86}Ta_{0.14})_3Sb_5$ with a superconducting transition temperature of 5 K [26]. We use a standard lock-in technique for dI/dV spectrum and a Pt-Ir tip as the non-superconducting tip [27, 28]. To obtain a superconducting tip, we intentionally drive the tip into the sample so that it picks up a portion of the superconducting sample material. The second sample is a single crystal of $KV_3Sb_5$, measured with an Nb tip, with a superconducting gap size of around 0.17 meV for $KV_3Sb_5$ and 1.33 meV for the tip [16]. When the junction normal-state resistance is small, one has to consider the effective resistance ($R_{series}$) in the circuit, which includes resistances from the bias cable, tip cable, and filter, etc. In practice, $R_{series}$ can be measured directly with a multimeter after bringing the tip into firm contact with the sample using the coarse approach with the piezo motor drive. In our instrument, $R_{series} \approx 5.3\ k\Omega$. For experiments with extremely small junction normal-state resistance, the voltage drop across the series resistance must be subtracted to obtain the actual junction bias $V_J = V_{Bias} - R_{series} \cdot I$ , where $V_{Bias}$ is the bias voltage supplied by the STM controller and $I$ is the tunneling current. The dI/dV values also have to be recalibrated according to $g_J \equiv dI/dV_J = dI/dV_{Bias}/(1 - dI/dV_{Bias} \cdot R_{series})$, since $V_J/I = V_{Bias}/I - R_{series}$. The detailed data-processing procedure is described in the Supplemental Material [29].

### III. RESULTS AND DISCUSSION

Fig. 1(a) shows the dI/dV spectrum acquired with a non-superconducting tip, which exhibits two clear and relatively sharp coherence peaks, corresponding to the sample superconducting gap magnitude of about $\Delta_s \approx 0.85\ meV$ (which is determined by the peak position). It is noted that the coherence-peak heights are not always symmetric, as is commonly observed in other samples in the literature [12, 30, 31]. When the tip is intentionally rendered superconducting, the spectrum at a normal-state resistance of $R_n \approx 4\ M\Omega$ also shows two sharp coherence peaks, as shown in Fig. 1(b); however, the peaks are located at higher energies around $\Delta_s + \Delta_t = 0.96\ meV$, which is the sum of the tip and sample superconducting gap magnitudes. This yields the tip superconducting gap $\Delta_t$ of about $0.11\ meV$. There are also some dip-hump features outside the main coherence peaks, which deserve further study to clarify their origin [31, 32].

Upon decreasing the junction normal-state resistance $R_n$, the Josephson coupling between the two superconductors increases. As a result, a non-zero and rapidly increasing ZBC appears in the dI/dV spectra, as shown in Fig. 2(b). The corresponding current-voltage (I-V) characteristics (Fig. 2(a)) show supercurrent appearing near zero bias. The current reaches a maximum $I_{max}$ before dropping.

As the junction resistance $R_n$ decreases further, the ZBC continues to increase, as shown in Fig. 2(d). Additionally, the overall dI/dV features start to change for the spectra with $R_n < R_0$ ($\sim 12.9\ k\Omega$) compared with those of junctions with higher $R_n$. The subgap states, apart from those around zero bias, develop in the spectra. When the junction normal-state resistance decreases to $R_0/2$ or less, the dI/dV features change substantially, i.e., the gap coherence peaks disappear. Also, ZBC starts to saturate and the conductance around zero bias is nearly flat over a finite bias range for a given spectrum, as shown in Figs. 2(e) and (f), which will be discussed in detail later.

In order to have a clearer view of the relationship between the ZBC and the state of the junction, we have plotted the ZBC versus the normal-state conductance $G_n = 1/R_n$, as shown in Fig. 3(a). We find that the ZBC versus $G_n$ curve can be roughly divided into three regimes: in regime 1 when $G_n \leq G_0 = 1/R_0$, $ZBC$ increases almost quadratically with $G_n$, with $ZBC \approx G_n^{1.892}$ from the fitting using $ZBC \approx G_n^m$ (where $m$ is the parameter to be determined from fitting), as also shown in the zoom-in view of Fig. 3(b), which is consistent with other studies [8]; in regime 2 where $G_0 \leq G_n \leq 2G_0$, ZBC deviates from the quadratic trend and increases more rapidly, possibly related to multiple Andreev reflections. The multiple Andreev reflections are also evidenced by the normalized zero-bias conductance,

ZBC/$G_n$, which approaches 2 when $G_n > G_0$, as expected from the Blonder-Tinkham-Klapwijk (BTK) theory [33, 34]. The microscopic mechanism for the high ZBC in this regime still requires further investigation; and finally in regime 3 when $G_n > 2G_0$, the ZBC increases much more slowly with the increase of $G_n$ and ultimately the ZBC appears to saturate, i.e., reaching a constant value within the experimental limit. This is especially puzzling at first sight and could easily (and mistakenly) be attributed to some exotic new physical phenomenon. A detailed analysis reveals, however, that this saturation is actually due to the existence of the resistors in series with the junction. In fact, the dI/dV directly measured from the lock-in amplifier is a combination of both the junction and an effective resistor (series resistance from cabling, etc.), i.e., $g_{Lock\ in} = 1/(g_J^{-1} + R_{series}) \leq 1/R_{series}$. As the normal-state conductance of the junction reaches a certain limit, $R_{series}$ would dominate. This would result in a saturation in the zero-bias conductance in the directly-measured dI/dV spectrum. This clearly shows that this is a measurement-induced saturation that in principle cannot be avoided. This also serves as a cautious reminder when interpreting zero-bias conductance saturation or quantization in studies of exotic physics like that of Majorana zero modes if the tip-sample junction resistance is extremely small. In addition, as one also notices from Fig. 2(f), the flat saturation regime has a finite width, which increases further as the normal-state conductance increases beyond the onset of saturation. This is likely related to the critical current of the junction, though the exact mechanism is not well understood yet.

The transition from tunneling to the point-contact regime is also supported by the current-distance (tip-sample relative distance) (*I-z*) spectroscopy, as shown in Fig. 3(d), where a clear change in slope is observed [35-39]. This is also discussed in more detail later. An important question is whether the tip and/or sample is stable during these high normal-state conductance experiments. We would also like to note that the tip and/or sample may be modified or remain intact. In Figs. 3(e) and 3(f), we show an example in which the Cs-rich surface of $Cs(V_{0.86}Ta_{0.14})_3Sb_5$ imaged after taking the dI/dV spectra has an additional hump, indicating that the tip and/or sample was somehow modified. The sample surface modification in STM has also been discussed in Refs [40-43]. However, all the features discussed above in the dI/dV spectra were still repeatable even after these modifications, showing that the features are intrinsic to the junction, rather than resulting from the modification process.

Having developed a deeper understanding of the Josephson current in JSTM, we now explore its key applications in the study of superconductivity. In the past few years, JSTM has emerged as a powerful technique as an atomic-scale probe for detecting the local superconducting order parameter and its spatial variations, including PDW states. For such a probe to work properly, both the tip and sample surface have to stay intact. In addition, as we have shown, the tip-sample junction can be in tunneling, transition, or point-contact regime. Therefore, it is crucial to identify the optimal regime in which JSTM can operate as such a probe. In order to explore this, we use a single crystal of $KV_3Sb_5$ and a Nb superconducting tip. Fig. 4(a) shows a spectrum of current *I* (left axis) versus relative tip-sample distance and the corresponding normal-state conductance (right axis). The *I-z* spectrum covers a much wider range compared to that in Fig. 3(d) and exhibits three distinct regimes with different slopes: tunneling (below about 10 nA), transition (roughly from 10 nA to 100 nA), and point-contact (above around 100 nA). The onset of the point-contact regime is evidenced by a much slower increase of current with the decrease of z, which corresponds to the normal-state conductance of about $0.5G_0 = e^2/h$.

The Cooper-pair density can be related to ZBC through $\rho_S(\boldsymbol{r}) \propto (I_C R_N)^2 \propto g(\boldsymbol{r},0)R_N^2(\boldsymbol{r}) = ZBC \cdot R_N^2(\boldsymbol{r})$, where $g(\boldsymbol{r},0) = ZBC$ is the zero-bias conductance at position $\boldsymbol{r}$ [14, 16, 44]. In Fig. 4b, we show the evolution of ZBC versus $G_N^2(\boldsymbol{r})$ for a wide range of $G_N$, corresponding to the data shown in Fig. 4a. Apparently, there are different portions of the curve that exhibit a linear relationship. An interesting question in the point-contact regime is whether there are any other changes (e.g., changes in bond length) at the tip and/or sample that could lead to the observed non-linearity in the high-$G_N$ regime and this deserves further study in the future. The mapping of $\rho_S(\boldsymbol{r})$ can be performed in constant-current mode or constant-height mode, as described by Liu et al. [14]. If the superconducting gap size of both the tip and sample is large enough [12, 13, 45], the PDW mapping can be performed with a relatively small normal-state conductance in constant-current or constant-height mode. However, if the sample or the tip superconducting gap is rather small, as in the case of $KV_3Sb_5$, the mapping of PDW is better performed in constant-height mode, as it takes much less time than constant-current mode.

The superconducting order parameter may be written as $\Delta(\boldsymbol{r}) = \Delta_0(\boldsymbol{r}) + \Delta_{\mathrm{P}}(\boldsymbol{r})$, where $\Delta_0(\boldsymbol{r})$ is the homogeneous component and $\Delta_{\mathrm{P}}(\boldsymbol{r})$ is the PDW order parameter. Ideally, in the constant-height mode, if the sample surface is perfectly flat and there is no drift in vertical (perpendicular to the sample surface) direction, mapping of $\Delta(\boldsymbol{r})$ at different positions corresponds to the same normal-state conductance $G_n$, i.e., the same vertical point in Fig. 4b. However, the sample surface is often not perfectly flat and the drift in vertical direction cannot always be avoided. This will introduce extra variation into $\Delta_0(\boldsymbol{r})$. For example, $\Delta_0(\boldsymbol{r_1})$ obtained at position $\boldsymbol{r_1}$ will be different from $\Delta_0(\boldsymbol{r_2})$ obtained at position $\boldsymbol{r_2}$. This extra variation in the homogeneous component of $\Delta_0(\boldsymbol{r})$ may introduce extra spots in the Fourier transform of $\Delta(\boldsymbol{r})$, making the analysis potentially more complicated. However, if the mapping of $\Delta(\boldsymbol{r})$ is performed in the linear regime of ZBC-$G_N^2(\boldsymbol{r})$ curve, then even if $G_N^2(\boldsymbol{r})$ varies due to non-flat sample surface or drift, as long as the actual $G_N^2(\boldsymbol{r})$ at different locations remains within the linear portion of ZBC-$G_N^2(\boldsymbol{r})$ curve (as shown by the fit in Fig. 4(b)), the impact of this extra variation in $\Delta_0(\boldsymbol{r})$ can be avoided to a large extent. Finally, as a practical technique to map PDW, one has to make sure the tip and sample are both stable and there are no changes even at atomic scale after the mapping. As an example, we show such a case on the Sb-terminated surface in $KV_3Sb_5$ in Fig. 4c, in which there is no noticeable change even at the atomic scale after the current versus z spectroscopy, with a bias of 4 meV and the current reaching as high as 320 nA. This requires high stability for both the tip and the sample.

## IV. CONCLUSIONS

In conclusion, we have investigated the Josephson current in JSTM from the tunneling regime to the deep point-contact regime. Upon entering the point-contact regime, the quadratic dependence of the zero-bias conductance on the normal-state conductance no longer holds. In addition, as it is pushed into the deep point-contact regime, the zero-bias conductance reaches a plateau. However, this plateau is found to result from the series resistance in the circuit. We further demonstrate that JSTM can only be used as a reliable atomic-scale probe for variation of the superconducting order parameter within certain ranges of normal-state conductance.

## ACKNOWLEDGMENTS

We thank D. Wu for sample preparation. We acknowledge the support from the National Key R&D Program of China (Grants No. 2023YFA1407300 and No. 2025YFA1411500), the National Natural Science Foundation of China (Grants No. 12374060, No. 12474153, No. 11804163), Guangdong Provincial Quantum Science Strategic Initiative (Grants No. GDZX2401001, No. GDZX2501002) and the Guangdong Natural Science Foundation (Grant No. 2026A1515010666). H.Q. acknowledges the start-up fund (Funding number: QD2301007) from Quantum Science Center of Guangdong-Hong Kong-Macao Greater Bay Area. H.D. acknowledges the support from the Young Scientists Fund of National Natural Science Foundation of China (Funding No. 12504162).

[1] B. D. Josephson, Possible new effects in superconductive tunnelling, Phys. Lett. 1, 251-253 (1962).

[2] B. D. Josephson, Coupled superconductors, Rev. Mod. Phys. 36, 216-220 (1964).

[3] B. D. Josephson, Supercurrents through barriers, Adv. Phys. 14, 419-451 (1965).

[4] P. W. Anderson and J. M. Rowell, Probable observation of the Josephson superconducting tunneling effect, Phys. Rev. Lett. 10, 230-232 (1963).

[5] A. A. Golubov, M. Yu. Kupriyanov, and E. Il'ichev, The current-phase relation in Josephson junctions, Rev. Mod. Phys. 76, 411-469 (2004).

[6] J. Šmakov, I. Martin, and A. V. Balatsky, Josephson scanning tunneling microscopy, Phys. Rev. B 64, 212506 (2001).

[7] O. Naaman, W. Teizer, and R. C. Dynes, Fluctuation dominated Josephson tunneling with a scanning tunneling microscope, Phys. Rev. Lett. 87, 097004 (2001).

[8] H. Kimura, R. P. Barber, Jr., S. Ono, Y. Ando, and R. C. Dynes, Josephson scanning tunneling microscopy: A local and direct probe of the superconducting order parameter, Phys. Rev. B 80, 144506 (2009).

[9] J. G. Rodrigo, H. Suderow, and S. Vieira, On the use of STM superconducting tips at very low temperatures, Eur. Phys. J. B 40, 483-488 (2004).

[10] J. G. Rodrigo and S. Vieira, STM study of multiband superconductivity in $NbSe_2$ using a superconducting tip, Physica C 404, 306-310 (2004).

[11] M. Graham and D. K. Morr, Josephson scanning tunneling spectroscopy in $d_{x^2-y^2}$-wave superconductors: A probe for the nature of the pseudogap in the cuprate superconductors, Phys. Rev. Lett. 123, 017001 (2019).

[12] M. H. Hamidian, S. D. Edkins, S. H. Joo, A. Kostin, H. Eisaki, S. Uchida, M. J. Lawler, E.-A. Kim, A. P. Mackenzie, K. Fujita, et al., Detection of a Cooper-pair density wave in $Bi_2Sr_2CaCu_2O_8$+x, Nature 532, 343-347 (2016).

[13] Z. Du, H. Li, S. H. Joo, E. P. Donoway, J. Lee, J. C. Seamus Davis, G. Gu, P. D. Johnson, and K. Fujita, Imaging the energy gap modulations of the cuprate pair-density-wave state, Nature 580, 65-70 (2020).

[14] X. Liu, Y. X. Chong, R. Sharma, and J. C. Seamus Davis, Discovery of a Cooper-pair density wave state in a transition-metal dichalcogenide, Science 372, 1447-1452 (2021).

[15] Q. Gu, J. P. Carroll, S. Wang, S. Ran, C. Broyles, H. Siddiquee, N. P. Butch, S. R. Saha, J. Paglione, J. C. Seamus Davis, et al., Detection of a pair density wave state in $UTe_2$, Nature 618, 921-927 (2023).

[16] H. Deng, H. Qin, G. Liu, T. Yang, R. Fu, Z. Zhang, X. Wu, Z. Wang, Y. Shi, J. Liu, et al., Chiral kagome superconductivity modulations with residual Fermi arcs, Nature 632, 775-781 (2024).

[17] X.-Y. Yan, H. Deng, T. Yang, G. Liu, W. Song, H. Miao, Z. Tu, H. Lei, S. Wang, B.-C. Lin, et al., Chiral pair density waves with residual Fermi arcs in $RbV_3Sb_5$, Chin. Phys. Lett. 41, 097401 (2024).

[18] M. Graham and D. K. Morr, Imaging the spatial form of a superconducting order parameter via Josephson scanning tunneling spectroscopy, Phys. Rev. B 96, 184501 (2017).

[19] M. Yao et al., Self-consistent theory of 2×2 pair density waves in kagome superconductors, Phys. Rev. B 111, 094505 (2025).

[20] W. Song et al., Switchable chiral pair density wave in pure $CsV_3Sb_5$, Phys. Rev. B 113, 125127 (2026).

[21] M. T. Randeria, B. E. Feldman, I. K. Drozdov, and A. Yazdani, Scanning Josephson spectroscopy on the atomic scale, Phys. Rev. B 93, 161115 (2016).

[22] Yu. M. Ivanchenko and L. A. Zil'berman, The Josephson effect in small tunnel contacts, Zh. Eksp. Teor. Fiz. 55, 2395 (1968); Sov. Phys. JETP 28, 1272 (1969).

[23] B. Jäck, M. Eltschka, M. Assig, M. Etzkorn, C. R. Ast, and K. Kern, Critical Josephson current in the dynamical Coulomb blockade regime, Phys. Rev. B 93, 020504 (2016).

[24] C. R. Ast, B. Jäck, J. Senkpiel, M. Eltschka, M. Etzkorn, J. Ankerhold, and K. Kern, Sensing the quantum limit in scanning tunnelling spectroscopy, Nat. Commun. 7, 13009 (2016).

[25] J. Senkpiel, S. Dambach, M. Etzkorn, R. Drost, C. Padurariu, B. Kubala, W. Belzig, A. Levy Yeyati, J. C. Cuevas, J. Ankerhold, et al., Single channel Josephson effect in a high transmission atomic contact, Commun. Phys. 3, 131 (2020).

[26] H. Deng, G. Liu, Z. Guguchia, T. Yang, J. Liu, Z. Wang, Y. Xie, S. Shao, H. Ma, W. Liège, et al., Evidence for time-reversal symmetry-breaking kagome superconductivity, Nat. Mater. 23, 1639-1644 (2024).

[27] G. Liu et al., Perspective: Imaging atomic step geometry to determine surface terminations of kagome materials and beyond, Quantum Front. 3, 19 (2024).

[28] W. Song et al., Many-body electronic structure in the pyrochlore superconductor $CsBi_2$ and spin-liquid candidate $Pr_2Ir_2O_7$, Phys. Rev. B 112, 245131 (2025).

[29] See Supplemental Material at [URL] for a description of how the raw dI/dV data are processed; another set of I-V and dI/dV spectra to show the saturation of ZBC with decreasing $R_n$.

[30] C.-L. Song, Y.-L. Wang, P. Cheng, Y.-P. Jiang, W. Li, T. Zhang, Z. Li, K. He, L. Wang, J.-F. Jia, et al., Direct observation of nodes and twofold symmetry in FeSe superconductor, Science 332, 1410-1413 (2011).

[31] B. Hu, H. Chen, Y. Ye, Z. Huang, X. Han, Z. Zhao, H. Xiao, X. Lin, H. Yang, Z. Wang, et al., Evidence of a distinct collective mode in kagome superconductors, Nat. Commun. 15, 6109 (2024).

[32] C.-L. Song, Y.-L. Wang, Y.-P. Jiang, Z. Li, L. Wang, K. He, X. Chen, J. E. Hoffman, X.-C. Ma, and Q.-K. Xue, Imaging the electron-boson coupling in superconducting FeSe films using a scanning tunneling microscope, Phys. Rev. Lett. 112, 057002 (2014).

[33] G. E. Blonder, M. Tinkham, and T. M. Klapwijk, Transition from metallic to tunneling regimes in superconducting microconstrictions: Excess current, charge imbalance, and supercurrent conversion, Phys. Rev. B 25, 4515-4532 (1982).

[34] J. Brand, P. Ribeiro, N. Néel, S. Kirchner, and J. Kröger, Impact of atomic-scale contact geometry on Andreev reflection, Phys. Rev. Lett. 118, 107001 (2017).

[35] N. Agraït, J. G. Rodrigo, and S. Vieira, Transition from the tunneling regime to point contact and proximity-induced Josephson effect in lead-normal-metal nanojunctions, Phys. Rev. B 46, 5814-5817 (1992).

[36] J. I. Pascual, J. Méndez, J. Gómez-Herrero, A. M. Baró, N. García, and Vu Thien Binh, Quantum contact in gold nanostructures by scanning tunneling microscopy, Phys. Rev. Lett. 71, 1852-1855 (1993).

[37] E. Tartaglini, T. G. A. Verhagen, F. Galli, M. L. Trouwborst, R. Müller, T. Shiota, J. Aarts, and J. M. van Ruitenbeek, New directions in point-contact spectroscopy based on scanning tunneling microscopy techniques (Review Article), Low Temp. Phys. 39, 189-198 (2013).

[38] J. Kröger, N. Néel, and L. Limot, Contact to single atoms and molecules with the tip of a scanning tunnelling microscope, J. Phys.: Condens. Matter 20, 223001 (2008).

[39] R. Berndt, J. Kröger, N. Néel, and G. Schull, Controlled single atom and single molecule contacts, Phys. Chem. Chem. Phys. 12, 1022-1032 (2010).

[40] L. Limot, J. Kröger, R. Berndt, A. García-Lekue, and W. A. Hofer, Atom transfer and single-adatom contacts, Phys. Rev. Lett. 94, 126102 (2005).

[41] J. Kröger, H. Jensen, and R. Berndt, Conductance of tip-surface and tip-atom junctions on Au(111) explored by a scanning tunnelling microscope, New J. Phys. 9, 153 (2007).

[42] J. Kröger, N. Néel, A. Sperl, Y. F. Wang, and R. Berndt, Single-atom contacts with a scanning tunnelling microscope, New J. Phys. 11, 125006 (2009).

[43] M. Ringger, H.-R. Hidber, R. Schlögl, P. Oelhafen, and H.-J. Güntherodt, Nanometer lithography with the scanning tunneling microscope, Appl. Phys. Lett. 46, 832-834 (1985).

[44] V. Ambegaokar and A. Baratoff, Tunneling between superconductors, Phys. Rev. Lett. 10, 486-489 (1963).

[45] W. Chen, W. Ren, N. Kennedy, M. H. Hamidian, S. Uchida, H. Eisaki, P. D. Johnson, S. O'Mahony, and J. C. Seamus Davis, Identification of a nematic pair density wave state in $Bi_2Sr_2CaCu_2O_8$+x, Proc. Natl. Acad. Sci. U.S.A. 119, e2206481119 (2022).

**Figures**

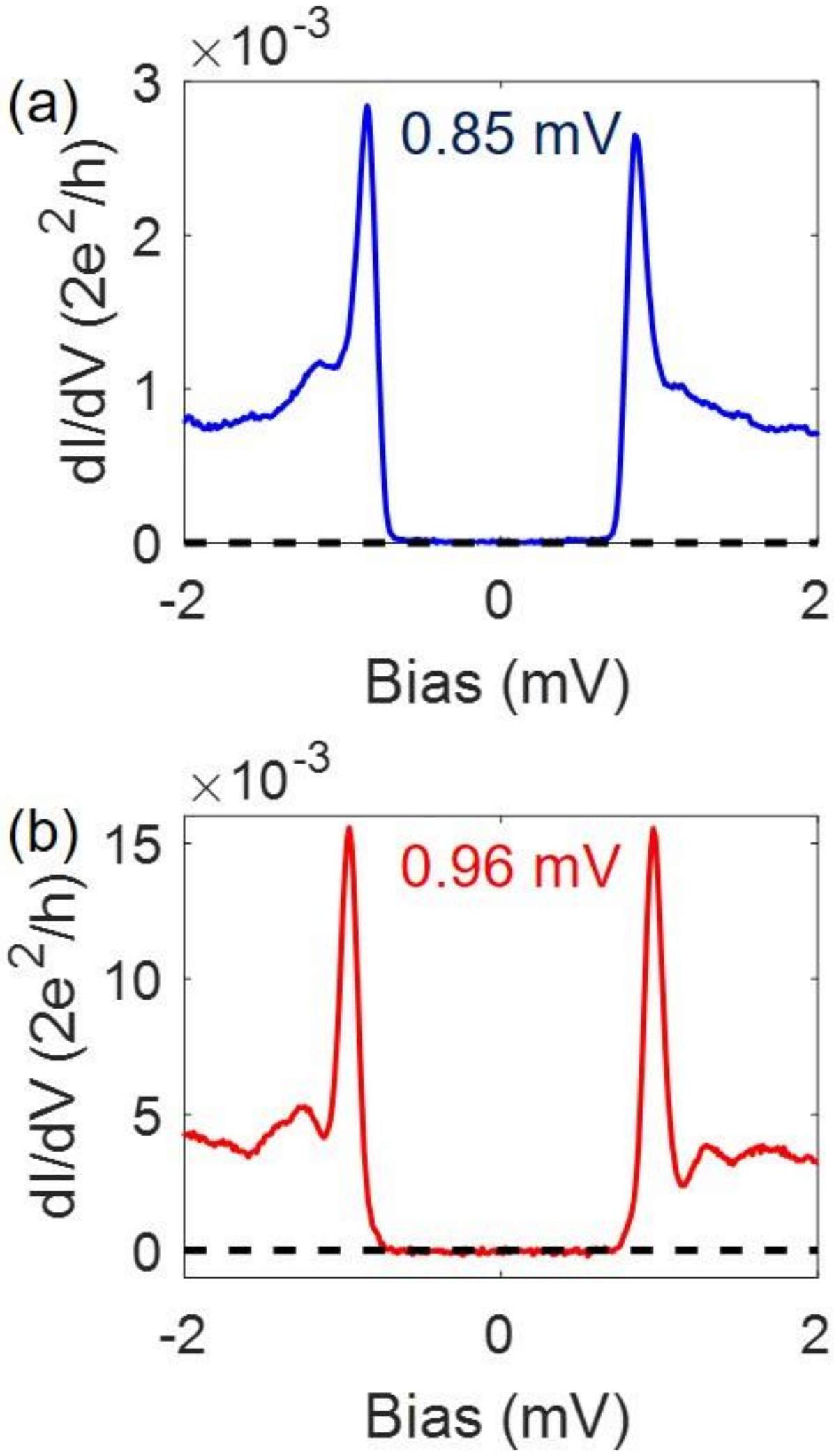

FIG. 1 dI/dV spectra acquired on the $Cs(V_{0.86}Ta_{0.14})_3Sb_5$ sample using a non-superconducting and a superconducting tip. (a) dI/dV spectrum using a non-superconducting tip ($R_n$ ~ 20 $M\Omega$). (b) dI/dV spectrum using a superconducting tip ($R_n$ ~ 4 $M\Omega$). Voltage drop on cables is neglected due to large normal-state resistance of the junction. The horizontal dotted lines are the respective zeros. Half of the distance between the two coherence peaks is about 0.85 mV and 0.96 mV, respectively, as indicated in the figures.

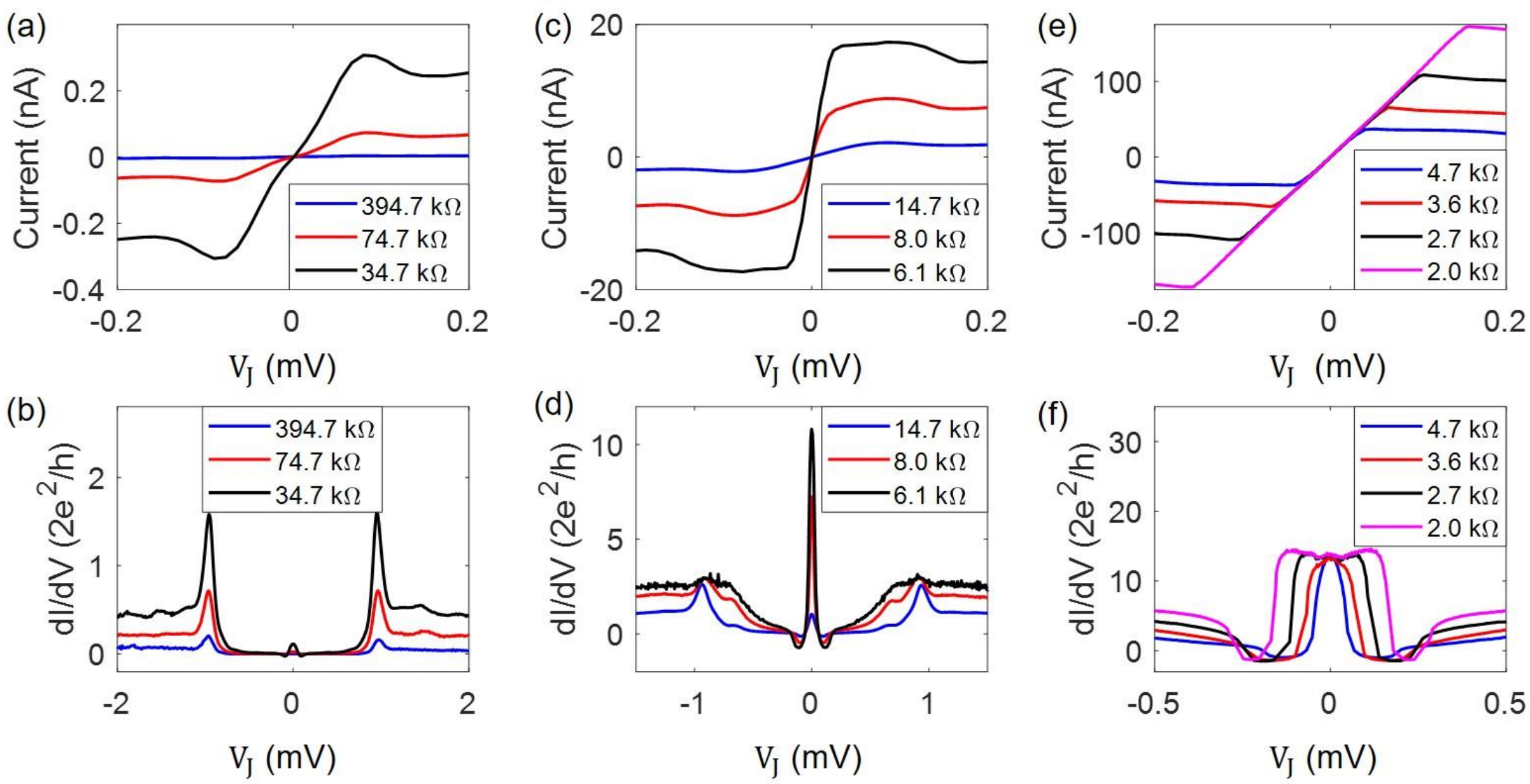

FIG. 2 I-V characteristics showing Cooper-pair current and corresponding dI/dV spectra acquired with a superconducting tip at smaller $R_n$. (a), (b) I-V and dI/dV spectra at $R_n$ = 394.7 kΩ, 74.7 kΩ, and 34.7 kΩ, respectively. (c), (d) I-V and dI/dV spectra at $R_n$ = 14.7 kΩ, 8.0 kΩ, and 6.1 kΩ, respectively. (e), (f) I-V and dI/dV spectra at $R_n$ = 4.7 kΩ, 3.6 kΩ, 2.7 kΩ, and 2.0 kΩ, respectively. Only data in small bias range are shown for some of the spectra for clarity.

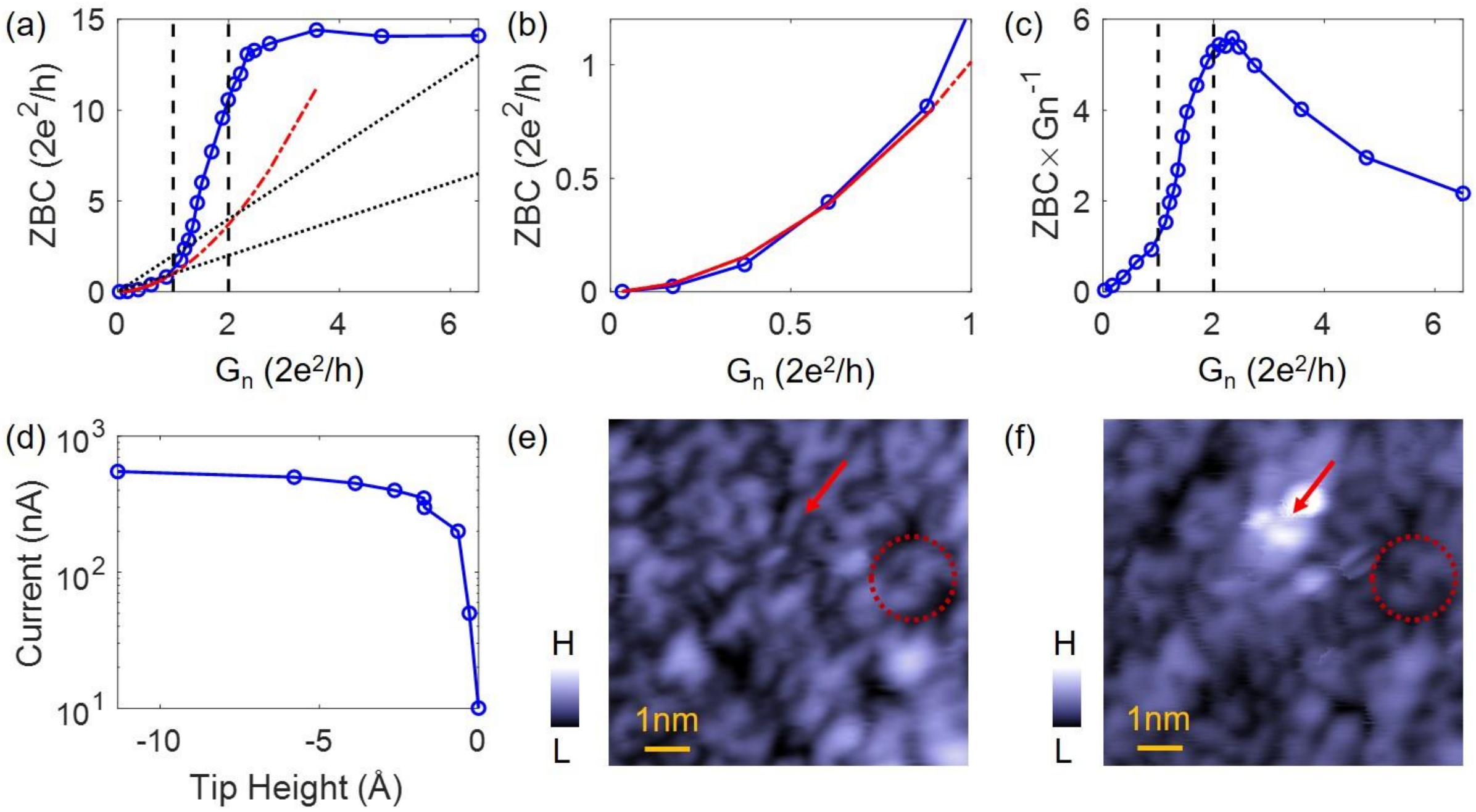

FIG. 3 ZBC versus junction normal-state conductance $G_n$ with fitting and surface modification. (a) The $ZBC$ versus $G_n$ extracted from dI/dV spectra. Two dotted lines corresponding to $ZBC = G_n$ and $ZBC = 2G_n$, respectively, are also shown for reference only. (b) Zoomed-in view of the part with $G_n < G_0$. (c) The $ZBC \times G_n^{-1}$ versus $G_n$ extracted from dI/dV spectra. (d) Tunneling current versus relative tip height during the dI/dV spectroscopy (note that the vertical axis is in logarithmic scale). (e) Topographic image before a series of dI/dV spectra. (f) Topographic image after a series of dI/dV spectra, showing clear surface modification. The red arrows indicate where the series of spectra were acquired. The dashed circles indicate a feature showing that (e) and (f) are actually the same area. H and L in the images indicate high and low, respectively.

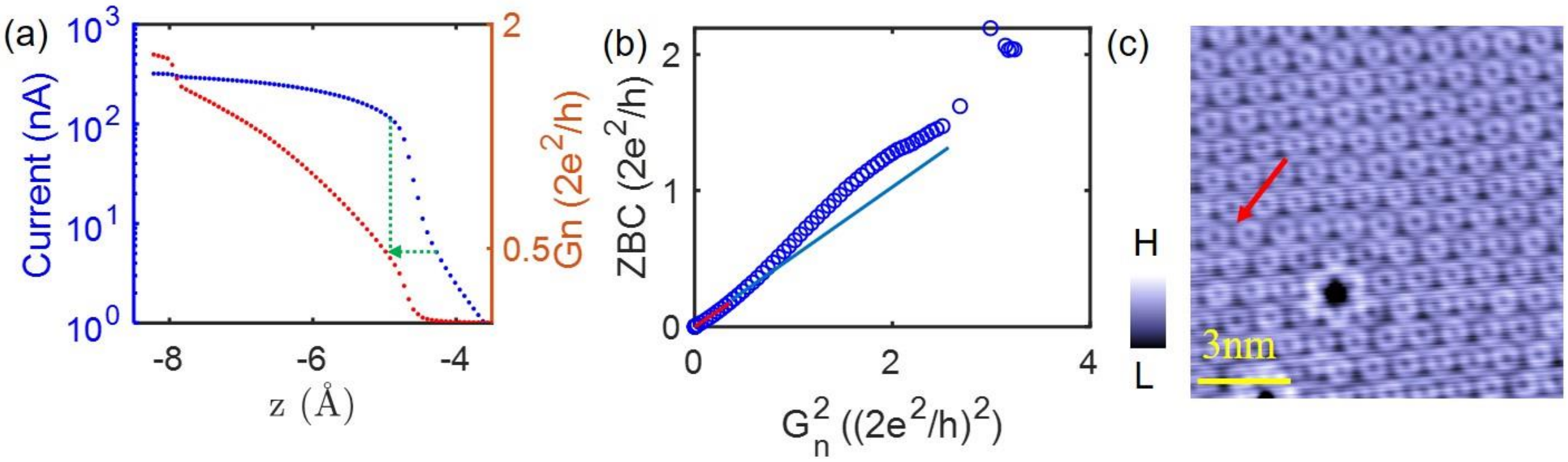

FIG. 4 Exploring an appropriate region for JSTM to be used as a probe for pair-density wave. (a) Tunneling current (left axis) and corresponding normal-state conductance (right axis) as a function of relative tip height z. The green arrow indicates the point where the normal-state conductance is around half of the conductance quantum and the vertical green dotted line indicates the corresponding point. (b) The zero-bias conductance ($ZBC$) versus square of normal-state conductance ($G_n^2$). The red line is a linear fit of the small $G_n^2$ region, while the light blue line is the extension of the red curve just to show the deviation of the linearity. (c) Topographic image (setpoint: V = 20 mV, $I_{Tunnel}$ = 1 nA) after a series of dI/dV spectra and I vs. z spectra, showing that there could be no surface modification even at atomic scale when the tip is very stable. The red arrow indicates where the series of spectra were acquired. H and L in the color bar indicate high and low in color scale, respectively.

# Supplemental Material for "Josephson spectroscopy study of kagome superconductors toward the deep point-contact regime"

Hailang Qin[1†], Xiao-Yu Yan[2], Hanbin Deng[2], Mu-Wei Gao[2], Guowei Liu[1], Yuanyuan Zhao[1], Jia-Xin Yin[1,2‡]

[1]Quantum Science Center of Guangdong-Hong Kong-Macao Greater Bay Area, Shenzhen, China.

[2]State Key Laboratory of Quantum Functional Materials, Department of Physics, and Guangdong Basic Research Center of Excellence for Quantum Science, Southern University of Science and Technology, Shenzhen 518055, China

[†]Contact author: qinhailang@quantumsc.cn

[‡]Contact author: yinjx@sustech.edu.cn

This Supplemental Material mainly includes the following:

- Another set of I-V and dI/dV spectra to show the saturation of ZBC with decreasing $R_n$.
- A description of how the raw dI/dV data are processed.

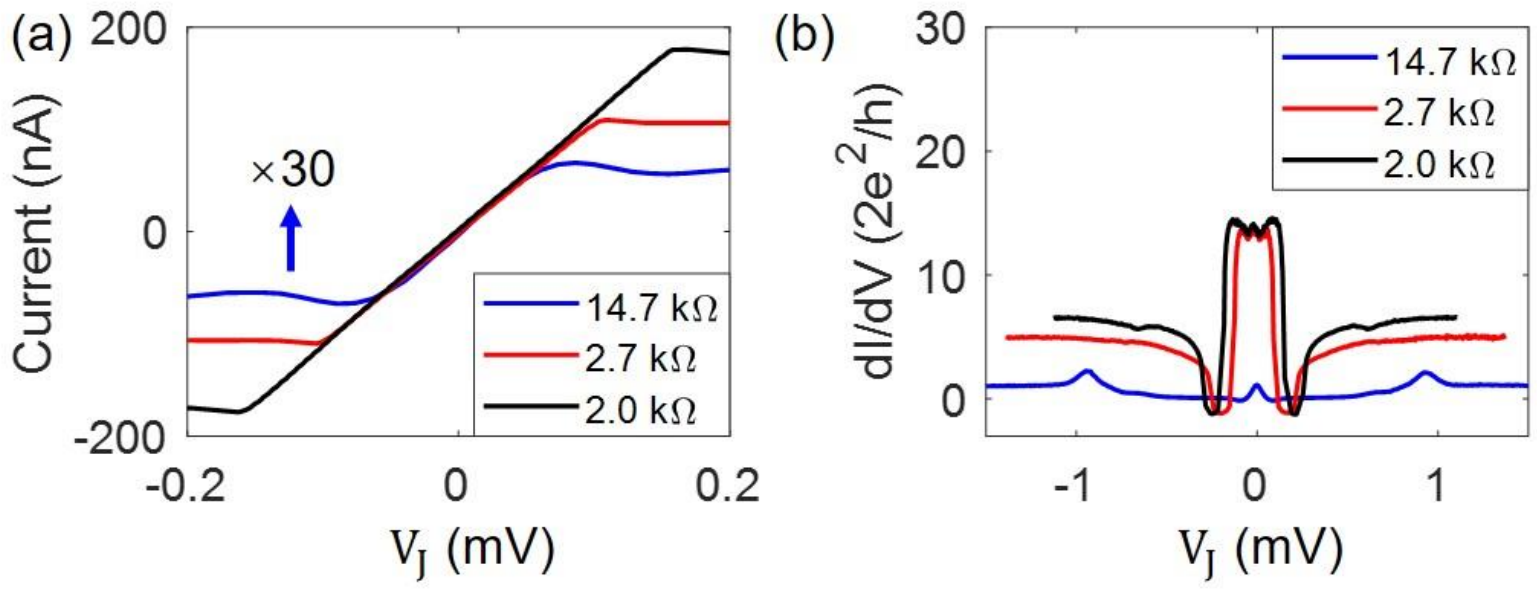

FIG. S1. Another set of data obtained on a $Cs(V_{0.86}Ta_{0.14})_3Sb_5$ sample at a different location, mainly to show the saturation behavior of the ZBC with decreasing $R_n$. (a) I-V. (b) dI/dV. Note that the I-V for $R_n = 14.7\ k\Omega$ has been multiplied by 30 for clarity.

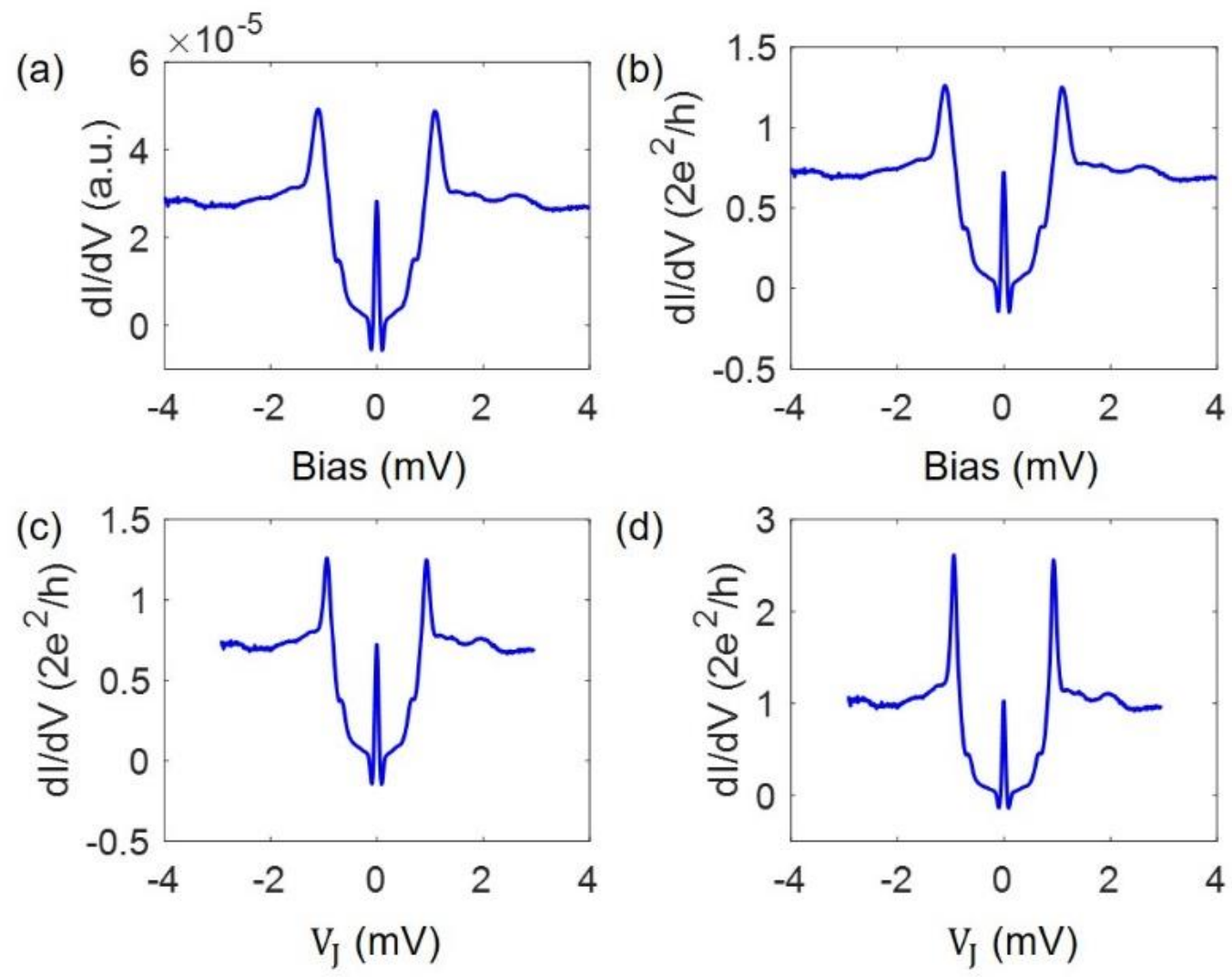

FIG. S2. An example illustrating how the raw dI/dV data are processed step by step. (a) The raw dI/dV spectrum as measured. Note that the unit for dI/dV is not determined yet at this stage. (b) The dI/dV values (vertical axis) in (a) have now been calibrated to match the numerical differentiation of measured I-V data (which has true units typically of nA/V or nS) so that the dI/dV values have real units. Here we have further converted the units into $2e^2/h.$ (c) Each point on the horizontal axis (bias) is calibrated by subtracting the voltage drop across the effective resistance $R_{series}$ using: $V_J = V_{Bias} - R_{series} \cdot I$ , where $V_{Bias}$ is the bias voltage supplied by the STM controller and $I$ is the tunneling current at each given bias. (d) The dI/dV value at each point is re-calibrated using: $g_J \equiv dI/dV_J = dI/dV_{Bias}/(1 - dI/dV_{Bias} \cdot R_{series})$, where $dI/dV_{Bias}$ is the value at each point in (c).